\documentclass[preprints,article,accept,pdftex,moreauthors]{Definitions/mdpi} 
\firstpage{1} 
\pubvolume{1}
\issuenum{1}
\articlenumber{0}
\pubyear{2025}
\copyrightyear{2025}
\datereceived{ } 
\daterevised{ } 
\dateaccepted{ } 
\datepublished{ } 
\hreflink{https://doi.org/} 

\usepackage{booktabs}  
\usepackage{array}     

\Title{Prenatal phthalate exposures and adiposity outcomes trajectories: a multivariate Bayesian factor regression approach}

\TitleCitation{Prenatal phthalate exposures and adiposity outcomes trajectories: a multivariate Bayesian factor regression approach}

\Author{Phuc H. Nguyen $^{1,}$*\orcidA{}, Stephanie M. Engel $^{2}$ and Amy H. Herring $^{3}$}

\AuthorNames{Phuc H. Nguyen, Stephanie M. Engel and Amy H. Herring}

\isAPAStyle{%
       \AuthorCitation{Nguyen, P., Engel, S., \& Herring, A.}
         }{%
        \isChicagoStyle{%
        \AuthorCitation{Nguyen, Phuc, Stephanie M. Engel and Amy H. Herring.}
        }{
        \AuthorCitation{Nguyen, P.; Engel, S.; Herring, A.}
        }
}

\address{%
$^{1}$ \quad LinkedIn Corporation, Sunnyvale, CA 94085, USA; phuc.nguyen.rcran@gmail.com\\
$^{2}$ \quad Department of Epidemiology, The University of North Carolina at Chapel Hill, Chapel Hill, NC 27599, USA; stephanie.engel@unc.edu\\
$^{3}$ \quad Department of Statistical Science, Duke University, Durham, NC 27708, USA; amy.herring@duke.edu}

\corres{Correspondence: phuc.nguyen.rcran@gmail.com}

\abstract{
Experimental animal evidence and a growing body of observational studies suggest that prenatal exposure to phthalates may be a risk factor for childhood obesity. Using data from the Mount Sinai Children’s Environmental Health Study (MSCEHS), which measured urinary phthalate metabolites (including MEP, MnBP, MiBP, MCPP, MBzP, MEHP, MEHHP, MEOHP, and MECPP) during the third trimester of pregnancy (between 25 and 40 weeks) of 382 mothers, we examined adiposity outcomes—body mass index (BMI), fat mass percentage, waist-to-hip ratio, and waist circumference—of 180 children between ages 4 and 9. Our aim was to assess the effects of prenatal exposure to phthalates on these adiposity outcomes, with potential time-varying and sex-specific effects. We applied a novel Bayesian multivariate factor regression (BMFR) that (1) represents phthalate mixtures as latent factors—a DEHP and a non-DEHP factor, (2) borrows information across highly correlated adiposity outcomes to improve estimation precision, (3) models potentially non-linear time-varying effects of the latent factors on adiposity outcomes, and (4) fully quantifies uncertainty using state-of-the-art prior specifications. The results show that in boys, at younger ages (4–6), all phthalate components are associated with lower adiposity outcomes; however, after age 7, they are associated with higher outcomes. In girls, there is no evidence of associations between phthalate factors and adiposity outcomes.}


\begin{document}



\section{Introduction}

The increasing prevalence of childhood obesity is a relevant public health concern. The 2017-2018 National Health and Nutrition Examination Surveys (NHANES) estimated that 19.3\% of children aged 2-19 years have obesity and another 16.1\% are overweight \citep{fryar2020cdc}. The prevalence of obesity among children and adolescents has increased more than three times in the last 30 years \citep{committee2012accelerating}. Obese children are at increased risk for several health conditions, including obesity in adulthood, type 2 diabetes, heart disease, arthritis, various cancers, and shorter life. Studies suggest that approximately 70\% obese children face a significant risk of heart disease in adulthood \citep{gollust2013framing}. Some estimates suggest that one-third of children born today (and half of Latino and black children) are expected to develop type 2 diabetes at some point in their lives \citep{narayan2003lifetime}. Childhood obesity also has negative implications for mental health, as it is associated with an increased risk of being bullied \citep{gollust2013framing}. 

Environmental chemical exposures, particularly exposure to phthalates, have been identified as risk factors for childhood obesity \citep{wu2020using, seo2022associations}, potentially by interfering with the body's endocrine system \citep{newbold2010impact, amato2021obesity}. Phthalates are a group of chemicals widely used in consumer products, such as toys, food packaging, or cosmetics, that are known to have endocrine disrupting properties \citep{hauser2005phthalates, encarnaccao2019endocrine}. In vivo and in vitro animal studies suggest that fetal development is a potential critical window of vulnerability to phthalate exposures, which may promote obesity, and the effect may depend on gender \citep{kim2014phth, national2016interplay, amato2021obesity}. Observational studies in humans, including birth cohort studies, further support these findings by showing associations between prenatal phthalate exposure and childhood adiposity outcomes \citep{engel2007prenatal, buckley2016prenatal, golestanzadeh2019association, berger2021prenatal}, some reporting sex-specific effects \citep{buckley2016prenatal, vafeiadi2018association, gao2022prenatal}. 

In this paper, we investigate the time-varying health effects of prenatal phthalate exposures on childhood obesity using data from the Mount Sinai Children’s Environmental Health Study (MSCEHS) \citep{engel2007prenatal}. The study measured the urinary phthalate metabolites of mothers during the third trimester of pregnancy and followed the health indices of children, including adiposity outcomes, between ages 4 and 9. We would like to identify groups of phthalate metabolites critical to these adiposity outcomes and estimate their time-varying effects and any sex-specific effects using a new modeling framework: the Bayesian multivariate factor regression model (BMFR). Our approach addresses the challenge of fully quantifying the uncertainty in estimating the time-varying effects of the mixture and provides several advantages tailored to our cohort data. 

Many studies have examined the time-varying effects of exposure to phthalates on adiposity outcomes or growth trajectories, but the results remain inconsistent \citep{botton2016phthalate, buckley2016prenatal, maresca2016prenatal, harley2017association, yang2018exposure, ye2021influence, maresca2016prenatal, heggeseth2019heterogeneity, gao2022prenatal, sol2025fetal}. A key limitation is that traditional statistical models, or even mixture methods not designed to estimate time-varying effects, lack full uncertainty quantification. In traditional statistical models, key modeling choices, such as the functional form of time-varying effects, are chosen or estimated first, then treated as fixed in the subsequent outcome model. For example, linear mixed-effects models or generalized estimating equations typically rely on interaction terms between a single exposure and age to model time-varying effects \citep{harley2017association, yang2018exposure, ye2021influence, sol2025fetal}. These approaches cannot consider all phthalate metabolites simultaneously due to multicollinearity, and assumptions about the functional form of time-varying relationships (e.g., linear or polynomial) are treated as fixed. Fixing these choices underestimates uncertainty, leading to confidence intervals that are too narrow and p-values that are artificially small. More advanced methods, such as growth mixture models, latent class growth models, or functional principal component analysis, can capture nonlinear growth trajectories, but require a second-stage model to estimate associations between mixture exposures and outcome trajectories \citep{gao2022prenatal, heggeseth2019heterogeneity}. In such multistage analyses, outputs from earlier stages (e.g., a number of selected trajectories or factors) are treated as fixed in later stages, again leading to underestimated uncertainty. Other advanced mixture methods, such as Bayesian kernel machine regression \citep{bobb2015bayesian}, quantile g-computation \citep{keil2020quantile}, weighted quantile sum regression \citep{czarnota2015assessment} do not model nonlinear time-varying effects. Bayesian varying-coefficient kernel machine regression (BVCKMR) is another advanced mixture method that can model nonlinear effects across exposure levels on outcome trajectories \citep{liu2018bayesian}. However, at fixed exposure levels, BVCKMR's estimated effects over time are limited to a functional form.

To address these limitations, BMFR applies state-of-the-art prior specifications to infer modeling decisions and fully quantify their uncertainty. The model includes variable selection priors for covariates, avoiding fixed subsets \citep{bai2018mbsp}; Gaussian process priors for flexible modeling of time-varying health effects without assuming a specific functional form \citep{bernardo1998regression}; and half-t priors for robust between-subject variances \citep{gelman2006prior}. To model the exposure mixture, BMFR assumes that structured variations in correlated exposures can be attributed to a small number of latent factors that also explain part of the variations in the outcomes, while fully quantifying the uncertainty in the number of factors through the multiplicative gamma process (MGP) prior \citep{dunson2011mgp}. This contrasts with principal component analysis (PCA) \citep{maresca2016prenatal}, which maximizes variation in exposures without regard to outcome relevance, or structural equation models \citep{hoyle1995structural}, which require a fixed number of factors.

Our implementation of BMFR also has custom features specific to our cohort. BMFR jointly models multiple adiposity outcomes, including BMI z-scores, waist circumference, waist-to-hip ration, and fat mass percentage, to improve estimation precision. Previous studies have generally examined these outcomes individually \citep{maresca2016prenatal, buckley2016prenatal, harley2017association, vafeiadi2018association, bowman2019phthalate, berger2021prenatal, ye2021influence}, hypothesized that mixtures affect all outcomes similarly \citep{maresca2016prenatal}, or searched for consistent results across outcomes \citep{harley2017association, berger2021prenatal, ye2021influence}. Because these outcomes are highly positively correlated and likely reflect shared mechanisms through which phthalates influence obesity, modeling them jointly is expected to improve estimation precision. Finally, BMFR quantifies uncertainty in imputing missing data and measurements below the limit of detection (LOD). Instead of relying on standard fixed-value imputations (e.g., mean or one-half of the LOD), BMFR handles imputation within the Markov Chain Monte Carlo (MCMC) process, propagating uncertainty in imputed values into the credible intervals for the effects of interest. An R package, optimized in C\texttt{++}, for BMFR is freely available at https://github.com/phuchonguyen/famr.
Section~\ref{sec:famr_data} describes in detail the data from the MSCEHS. Section~\ref{sec:famr_model} describes our proposed model BMFR and its prior specifications. Section~\ref{sec:famr_sims} validate BMFR's utility through simulation studies. Section~\ref{sec:famr_sinai} describes the analysis of time-varying health effects of prenatal phthalate exposures on childhood obesity using the MSCEHS data.

\section{Data}
\label{sec:famr_data}

\subsection{Study population}
\begin{figure}[H]
\includegraphics[trim={0 0 0 5cm}, clip, width=0.6\textwidth]{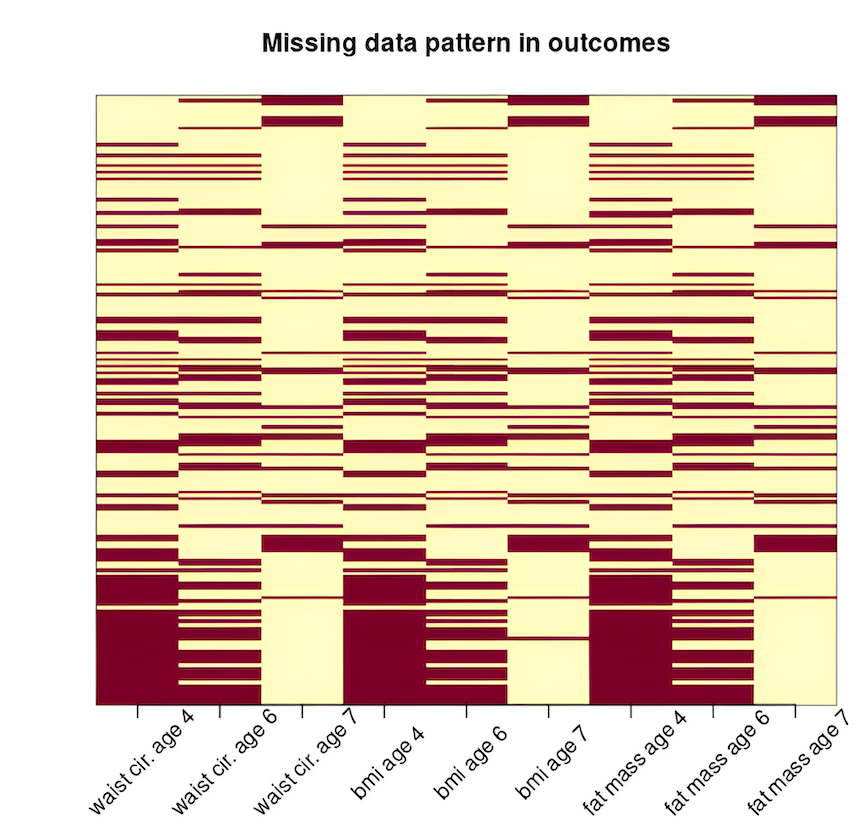}
\caption{Loss to follow-ups patterns for WC, BMIz, and FMP in the study population. Each row on the y-axis corresponds to one of 382 babies in MSCEHS (randomly ordered), and each column on the x-axis represents an outcome. Cells are shaded maroon if the value is observed and shown in light yellow if the value is missing.}
\label{fig:viz-outcomes}
\end{figure}

Between 1998 and 2002, MSCEHS recruited 479 first-time mothers with singleton pregnancies from the Mount Sinai Diagnostic and Treatment Center and two adjacent private practices in New York City. Among these women, 75 were excluded due to medical complications (n = 3), infant or fetal death (n = 2), very premature birth (before 32 weeks of gestation or <1,500 g; n = 5), miscarriage (n = 1), delivery of an infant with genetic abnormalities or malformations (n = 5), inability to obtain biologic samples before delivery (n = 12), relocation or transfer to a hospital outside of New York City (n = 28), or loss of follow-up (n = 19)\citep{engel2007prenatal}. The final cohort consisted of 404 babies with birth data recorded. The children were invited back for three follow-up visits at ages 4-5.5, 6, and 7-9. Of the 404 babies, 382 had their mothers' prenatal concentrations of phthalate metabolites measured in urine. Two additional observations were excluded because they had very dilute urine ($<10$ mg / dL creatinine) that can produce inaccurate biomarker measurements \citep{buckley2016prenatal}. Of the 382 babies, only 180 came back for at least one follow-up visit. This results in 362 total visits for fat mass percentage, 363 total visits for body mass index (BMI), 364 total visits for waist-to-hip ratio, and 364 total visits for waist circumference. Figure~\ref{fig:viz-outcomes} shows the pattern of loss to follow-up of these 382 babies. All observed outcome measurements were included in our analysis. Our analysis was a secondary analysis of the de-identified data from the MSCEHS study.

\unskip

\subsection{Phthalate exposures}
\begin{figure}
    \includegraphics[width=0.43\textwidth]{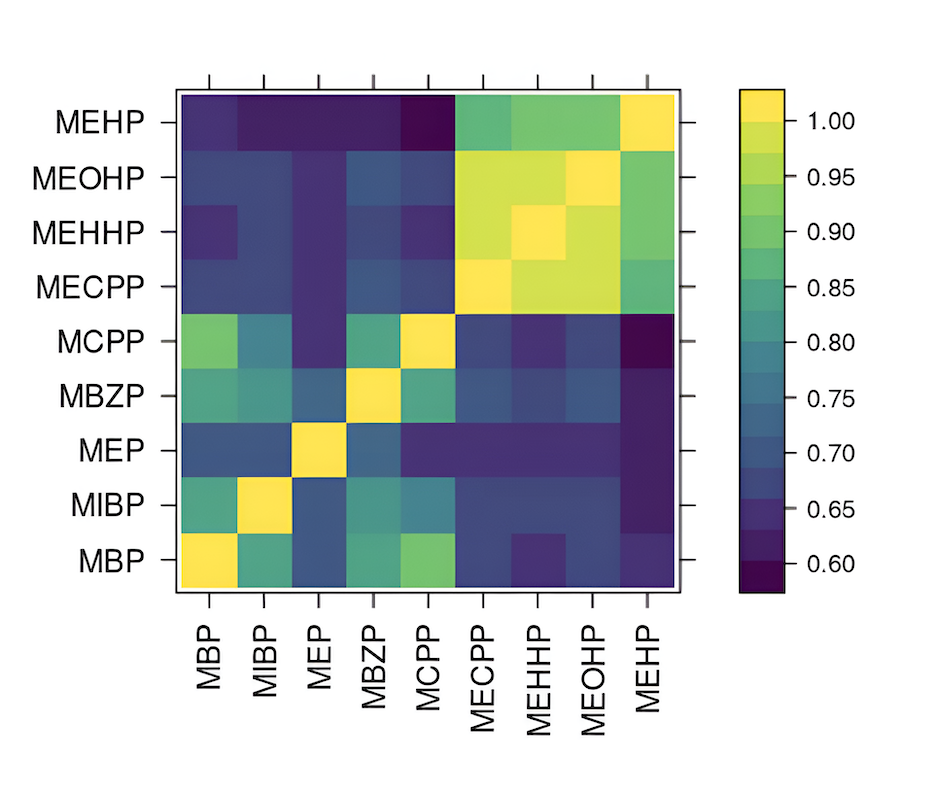}
    \includegraphics[width=0.47\textwidth]{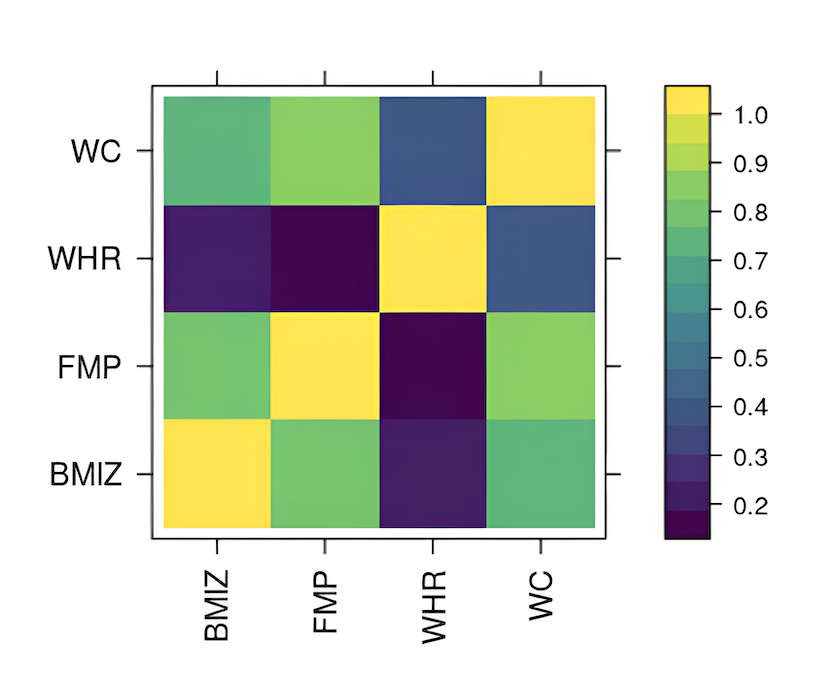}
    \caption{Left: Sample Pearson's correlation of phthalate metabolites concentrations. Right: Sample Pearson's correlation of adiposity outcomes.}
    \label{fig:viz-cor}
\end{figure}

Mothers who were pregnant between 25 and 40 weeks (with a mean of 31.5 weeks) provided a urine sample that was analyzed for various phthalate metabolites by the CDC laboratory, including MEP, MnBP, MiBP, MCPP, MBzP, MEHP, MEHHP, MEOHP, and MECPP. The DEHP group consists of MEHP, MEHHP, MEOHP, and MECPP. To account for inaccuracies in analytical standards, correction factors of 0.72 and 0.66 were applied to MBzP and MEP concentrations and limits of detection (LOD) respectively \citep{cdc2012}. Urinary concentration was measured using creatinine. To adjust for the dilution of urine samples, we standardize the metabolites' concentrations by a Cratio, as well as include creatinine concentration as a covariate in your analysis as suggested by \cite{o2017lipid}. The Cratio is calculated as the ratio between predicted creatinine  conditional on observed covariates of the mother including the mother's age, mother's BMI, mother's gestational weight gain adequacy category, mother's smoking status, mother's education, mother's race and observed creatinine \citep{o2017lipid}. As seen in Figure~\ref{fig:viz-cor} (Left), phthalate metabolites are highly and positively correlated with each other, especially those within the DEHP group and those within the non-DEHP group. There are 4 MiBP, 1 MEP, 1 MBzP, 4 MCPP, 1 MECPP, 1 MEHHP, 1 MEOHP, and 15 MEHP measurements under their respective LOD of detections in total. We impute these within the MCMC sampler as discussed later.

\unskip

\subsection{Adiposity outcomes}

The MSCEHS assessed the weight and body composition of the infants through bioelectrical impedance analysis using a pediatric Tanita scale at three follow-up visits scheduled at approximately 4 years, 6 years, and 7 years, though the age the children at actual visits ranges from 4 to 10 years (or 48 to 122 months). We consider the following four outcomes in our multivariate analysis: fat mass percentage (FMP), BMI z-score (BMIz), waist-to-hip ratio (WHR), and waist circumference (WC). FMP is based on fat mass estimates reported by the Tanita scale (model TBF-300; Tanita Corporation of America) and calculated as (fat mass/weight) $\times 100$. BMI is calculated as weight (in kilograms)/height (in meters)$^2$. It is then standardized by age and sex using a CDC SAS macro \citep{cdc2004} to produce BMIz. There are 362 total visits for FMP, 363 for BMIz, 364 for WHR, and 364 for WC from 180 subjects across all their follow-up visits. Figure~\ref{fig:viz-cor} (Right) shows that these outcomes (except WHR) are highly and positively correlated with each other.

\subsection{Covariates}
Mothers were interviewed for 2 hours during enrollment to gather covariate data. The computerized perinatal database at Mount Sinai Hospital was used to obtain pregnancy and delivery information. Gestational weight gain adequacy is calculated by dividing the observed gestational weight gain (last pregnancy weight minus self-reported pre-pregnancy weight) by the expected gestational weight gain based on the 2009 Institute of Medicine guidelines times 100 \citep{iomnrc2009}. We categorize gestational weight gain as inadequate if the ratio is $<86\%$, adequate if $86\% - 120\%$, and excessive if $>120\%$. The final baseline covariates include the mother's age, mother's BMI, mother's gestational weight gain adequacy category, mother's smoking status, mother's education, mother's race, whether mother breastfed, child's sex, child's birth weight, and creatinine concentration. The children's age in months was also recorded at each follow-up visit and included as a covariate. Table \ref{tab:sample-characteristics} summarizes the baseline covariates of children with at least one follow-up visit included in our analysis. Most covariates have no missing values or only negligible amounts, except for maternal gestational weight gain. The baseline characteristics of the male and female subsamples are similar, with only minor differences in race / ethnicity distribution.

\begin{table}[H]
\renewcommand{\arraystretch}{1.2} 
\resizebox{\textwidth}{!}{%
\begin{tabular}{lccc}
\toprule
\textbf{Characteristics} & \textbf{Study sample} & \textbf{Male sample} & \textbf{Female sample} \\
                         & n (\%)                & n (\%)               & n (\%) \\
                         & mean $\pm$ SD         & mean $\pm$ SD        & mean $\pm$ SD \\
\midrule
Total (n) & 180 & 97 & 83 \\

Race/ethnicity & & & \\
\hspace{2em} Non-Hispanic white & 33 (18.3) & 20 (20.6) & 13 (15.7) \\
\hspace{2em} Non-Hispanic black & 51 (28.3) & 27 (27.8) & 24 (28.9) \\
\hspace{2em} Hispanic or other & 96 (53.3) & 50 (51.5) & 46 (55.4) \\

Maternal age at delivery (years) & 24.4 $\pm$ 6.4 & 24.6 $\pm$ 6.7 & 24.2 $\pm$ 6.1 \\
Maternal education ($\geq$ college degree) & 39 (21.7) & 21 (21.6) & 18 (21.7) \\
Maternal prepregnancy BMI (kg/m$^2$) & 23.9 $\pm$ 4.7 & 24.0 $\pm$ 5.0 & 23.8 $\pm$ 4.3 \\
\hspace{2em} Missing & 0 & 0 & 1 (1.2) \\
Maternal gestational weight gain (lbs) & 40.8 $\pm$ 18.4 & 39.0 $\pm$ 18.0 & 42.8 $\pm$ 18.7 \\
\hspace{2em} Missing & 22 (12.2) & 13 (13.4) & 9 (10.8) \\
Maternal smoking during pregnancy & & & \\
\hspace{2em} Ever & 31 (17.3) & 18 (18.6) & 13 (15.7)\\
\hspace{2em} Never & 149 (82.7) & 79 (81.4) & 70 (84.3)\\
Breastfed & & & \\
\hspace{2em} Ever & 113 (62.8) & 58 (59.79) & 55 (66.3) \\
\hspace{2em} Never & 66 (36.7) & 39 (40.2) & 27 (32.5) \\
\hspace{2em} Missing & 1 (0.6) & 0 & 1 (1.2) \\
Child's birthweight (g) & 3296 $\pm$ 458 & 3352 $\pm$ 475 & 3229 $\pm$ 430 \\

\bottomrule
\end{tabular}
}
\caption{Sample characteristics of participants with at least one follow-up, stratified by child sex, in the Mount Sinai Children’s Environmental Health Study 1998–2002.}
\label{tab:sample-characteristics}
\end{table}

\section{Bayesian multivariate factor regression for time-varying effects}
\label{sec:famr_model}

\subsection{Model correlated chemical mixtures with a latent factor model}
Let $X_i$ be a $p$-vector of correlated metabolite concentrations measured during the third trimester of pregnancy for subject $i$. We assume that variation in $X_i$ can be attributed to $K < p$ latent variables:

\begin{align}
    X_i &\sim N_p(\Theta \eta_i , \Sigma_X) \\
    \eta_i &\sim N_K(0, \boldsymbol{I})\\
    \Sigma_X &= diag(\sigma_{X,1}^2, ..., \sigma_{X,p}^2)
\end{align}

\noindent where $\eta_i$ is a $K$-vector of unobserved latent factors of subject $i$, $\Theta$  is the factor loadings matrix, and $\sigma^2_{X,1}, ..., \sigma^2_{X,p}$ are idiosyncratic noise variances. We assume the exposures have been mean-centered and remove the intercepts. Independent priors on each $\sigma^2_{X,1}...\sigma^2_{X,p}$ are chosen to be those often used in factor analysis. 
We use the multiplicative gamma process (MGP) prior on the factor loadings to learn sparse loadings structure and infer the number of factors $K$ \citep{dunson2011mgp}:


\begin{align}
    \theta_{jk} &\sim N(0, \phi_{jk}^{-1} \tau_k^{-1})\\
    \phi_{jk} &\sim G(v/2, v/2), \quad j=1,...,p; k=1,...,K\\
    \tau_h &= \prod_{l=1}^h \delta_l, \quad \delta_1 \sim G(a_1, 1), 
    \quad \delta_{l \geq 2} \sim G(a_2, 1)
\end{align}

\subsection{Model correlated outcomes as a function of latent factors}

Let $Y_{it} = (y_{it1},...,y_{itq})^T$ be a vector of $q$ outcomes at follow-up time $t=1,..., T_i$, where $T_i$ is the number of follow-ups with at least one measured outcome for subject $i$. We assume each outcome at each follow-up has been mean-centered and remove the intercepts. We also assume the variation in the outcomes $Y_{it}$ can be decomposed into the variation explained by the latent factors of $X_i$, the variation due to unobserved factors and idiosyncratic noise:

\begin{align}
    Y_{it} &\sim N_q(B(t) \eta_i + \xi_i , \Sigma_Y)\\
    \xi_i &\sim N_H(0, \nu^2 \Sigma_Y) \\
    \Sigma_Y &\sim IW(s_0, S_0)
\end{align}

\noindent where $\nu^2$ has a half-t prior with a small degree of freedom for an uninformative prior that still behaves well in the case that between-subject variance $\nu^2$ is close to zero, as suggested by \cite{gelman2006prior}. Random variables $\xi_i$ are subject-level random intercepts. $\Sigma_Y$ is the residual covariance, which describes variances and covariances in the outcome due to unmeasured factors as well as random noise. We can interpret $\frac{\nu^2}{ \nu^2 + 1}$ as the proportion of total residual variation explained by between-subject variation, and $\frac{1}{ \nu^2 + 1}$ as the proportion of residual variation explained by within-subject variation.

\subsection{Model health effects as flexible functions of time}

$B$ is a $(q \times K)$ matrix of regression functions, which is of primary interest in our analysis. Element $jk$ in $B$ models the effect of the $k^{th}$ latent factor on the $j^{th}$ outcome that can vary smoothly and flexibly over follow-up times. We consider Gaussian processes as priors to learn these smooth regression functions on a discrete-time grid $t=1,...,T$, where $T$ is the total number of unique ages at which children had follow-up visits. At the same time, we want to incorporate our belief that effects across adiposity outcomes, which are driven by similar mechanisms through which phthalates interfere with the body's hormones, should be correlated. As a result, instead of placing independent Gaussian process priors on elements of $B$, we adopt the following factorization:

\begin{align}
    B(t) &= \Lambda U(t)\\
    u_{hk} &\sim GP(0, c_\kappa(t, t')), \quad c_\kappa(t, t') = e^{-\frac{1}{2} [\frac{(t - t')}{\kappa}]^2} \\
    \lambda_{jh} &\sim MGP, \quad j=1,...,p; h=1,...,H; k=1,...,K
\end{align}

\noindent where $\Lambda$ is a $q \times H$ matrix, with $H \leq K$, that linearly combines $H$ independent basis functions into elements of $B$. A similar factorization was used by \cite{fox2015bayesian}, but their work focused on covariance regression. We also use the MGP prior on the basis functions loadings to learn sparse loadings structure and help infer the number of basis functions $H$ \citep{dunson2011mgp}. $U$ is a matrix of independent basis functions. We choose the Gaussian kernel $c(t, t')$ for all elements of $U$ to ensure the time-varying effects are smooth functions of time with the same wiggliness encoded in a shared length scale $\kappa$. Since the input is on a grid, we place a uniform prior on a grid of plausible values for $\kappa$. Conditional on $\Lambda$, the $k^{th}$ column of $B$ has a separable Gaussian process prior:

\begin{align}
    B_k &\sim N_{q \times T}(0, \Lambda \Lambda^T, C)
\end{align}

\noindent where $\Lambda \Lambda^T$ describes the covariance in the regression functions of factor $k$ across outcomes, and $C_{rs} = c_\kappa(t_r, t_s)$. Thus, $Cov(B_{jk}) = [\Lambda \Lambda^T]_{jj} C$, so we set the amplitude of the kernel to 1 for identifiability.

\subsection{Model linear effects and interactions of covariates}

Let $Z_{it}$ be a $L$-vector of covariates, including both baseline covariates and those collected at follow-up time $t$. For covariates where the linear relationships are reasonable, we can add a linear effect term to equation (4) as follows:

\begin{align}
    Y_{it}&\sim N_q(B(t) \eta_i + B^{(c)}Z_{it} + \xi_i, \Sigma_Y)
\end{align}

We endow the regression coefficient matrix $B^{(c)}$ with a global-local shrinkage prior on matrix normal parameters of \cite{bai2018mbsp}:

\begin{align}
    B^{(c)} &\sim N_{q \times L}(0,\Sigma_Y, \Psi^{(c)}) \\
    \Psi^{(c)} &= diag(\psi_{1}^{(c)},...,\psi_{L}^{(c)}) \\
    \psi_{l}^{(c)}|\zeta_{l}^{(c)} &\sim G(u, \zeta_{l}^{(c)}) \\
    \zeta_{l}^{(c)} &\sim G(v, r)
\end{align}

Shrinkage parameter $\psi_{l}^{(c)}$ provides variable selection to determine if predictor $l$ is important to all outcomes, which fits our application. Outcome-specific effects within column $l$ can additionally shrink toward zero. The authors in \citep{bai2018mbsp} suggested setting the global shrinkage parameter $r = 1/(K\sqrt{n \ln{n}})$ to satisfy sufficient conditions for posterior consistency \citep{bai2018mbsp}. When $u=v=1/2$, this is the horseshoe prior \citep{tpb11, bai2018mbsp}. A similar setup can be used for any linear interactions between the latent factors and covariates.

\subsubsection{Imputation of censored and missing data}

We have very few missing outcomes at recorded follow-up visits (2 missing FMP and 1 missing BMIz measurement out of 364 observations). In the case that it's reasonable to assume outcomes are missing at random conditional on observed covariates and birth weight data \citep{buckley2016nyc}, we impute them during MCMC. We sample $Y_{it, mis}$ given $Y_{it, obs}, ~\boldsymbol{\omega}$ from a conditional multivariate normal, where $\boldsymbol{\omega}$ are all unknown parameters:

\begin{align}
    Y_{it, mis} &| Y_{it, obs}, ~\boldsymbol{\omega} \sim N(m, V)\\
    m &= B(t)_{mis} \eta_{i, mis} + \Sigma_{Y, mis, obs} \Sigma_{Y, obs, obs}^{-1} \Bigr[ Y_{it, obs} - B(t)_{obs}\eta_{i, obs} \Bigr]\\
    V &= \Sigma_{Y, mis, mis} - \Sigma_{Y, mis, obs} \Sigma_{Y, obs, obs}^{-1} \Sigma_{Y, obs, mis}
\end{align}

\noindent where $B(t)_{mis}, ~\eta_{i, mis}$ are parameters corresponding to the indices of the missing values, $B(t)_{obs}, ~\eta_{i, obs}$ are parameters at observed indices, $\Sigma_{Y, mis, obs}$ is the covariances between missing and observed indices, $\Sigma_{Y, mis, mis}$ is the covariance matrix of missing indices, and $\Sigma_{Y, obs, obs}$ is the covariance matrix of observed indices.

Moreover, we often observe censored metabolite concentrations that are below the limit of detection (LOD). The LOD is defined as the lowest concentration of an analyte in a sample that can be reliably distinguished from the highest concentration of the same analyte in a sample with no such analyte \citep{armbruster2008limit}. We can impute metabolite concentrations under the LOD by sampling from a conditional truncated normal at each MCMC iteration:

\begin{equation}
    X_{ij} | X_{ij} \in \{-\infty, \log(LOD_j) \}, \boldsymbol{\omega} \sim TN(\theta_{j.}^{T}\eta_i, \sigma^{2}_{X,j}, -\infty, \log(LOD_j))
\end{equation}

\noindent where $LOD_j$ is the LOD of the $j^{th}$ chemical, $\theta_{j.}$ is the $j^{th}$ row of loading matrix $\Theta$, and $TN(m, v, a, b)$ is a truncated normal distribution with mean $m$, variance $v$, and support $[a, b]$.

\subsection{Posterior computation}

See Appendix for full conditional and adaptive Metropolis-within-Gibbs updates.

\section{Simulations}
\label{sec:famr_sims}

We compare our proposed method BMFR with the following approaches: 1) two-stage univariate regressions, 2) BVCKMR \citep{liu2018bayesian}, and 3) a baseline mean model. In the two-stage approach, we reduce the dimension of the exposures using PCA, keeping the first few principal components (PCs), and then fit LMM for each outcome separately. The baseline mean model returns the mean of each outcome. We simulate data from the following three scenarios to demonstrate our method's utility compared to existing approaches. For all scenarios, we generate data for $n=200$ subjects. We generate the exposures according to a factor model, creating correlated exposures with group structures, similar to the observed phthalates metabolites:

\begin{align}
    X_i &= \Theta \eta_i + e_i \quad \text{for } i=1,...,n \\
    \eta_i &\sim N_{K^\ast}(0, I), \quad e_i \sim N_p(0, I)\\
\end{align}

We generate sparse $\Theta$ so that every five metabolites load onto one factor for $p=10$ and $K^\ast = 2$ latent factors. The non-zero entries of $\Theta$ are sampled from $N(0, 1)$. We generate $q=5$ outcomes measured at $T_i=10$ time points for all subjects:

\begin{align}
    Y_{it} &= \boldsymbol{g}(X_i, t) + \xi_i + \epsilon_{it}\\
    \xi_i &\sim N_q(0, C_\xi), \quad \epsilon_{it} \sim N_q(0, C_\epsilon)
\end{align}

We use different exposure-response functions $\boldsymbol{g}$ including time-varying effects, different distributions for the random intercept $\xi_i$ and random error $\epsilon_{it}$ for each scenario. Below is the description of each simulation scenario:

\begin{enumerate}
    \item Linear exposure-response function where the first two factors are important, independent responses:
    \begin{align}
        \boldsymbol{g}(X, t) &= \boldsymbol{h}(\eta, t) = \beta_1^T \eta + \beta_2^T \eta t, \quad j=1,...,10; k=1,...,K^\ast \\
        \beta_{1jk} &\sim  U(-3, 3), \quad \beta_{2jk} \sim U(-0.5, 0.5)\\
        C_\xi &= I, \quad  C_\epsilon = 0.5 I
    \end{align}
    \noindent where $\boldsymbol{h}$ is the latent factor-response function. The induced effects of $X$ range from 0 to 1. Under this setting, all assumptions for PCA-LMM are satisfied, though it does not propagate the uncertainty from the first stage.
    
    \item Non-linear time-varying exposure-response function where the first two factors are important, positively correlated responses:
    \begin{align}
        \boldsymbol{g}(X, t) &= \boldsymbol{h}(\eta, t) = \left[ \beta u(t)^T \right] \eta\\
        u_1(t) &= 3.5 / (1 + exp(-3t + 25)), \quad u_2(t) = 9 dnorm((t - 5.5)/1.5) \\
        \beta &\sim N_q(0, I)
    \end{align}
    \noindent where $dnorm(t)$ is the standard normal pdf at $t$. Covariances $C_\xi$ and $C_\epsilon$ have composite symmetry structure with high correlation of 0.7. Under this setting, all assumptions for our BMFR model are satisfied.
    
    \item Quadratic exposure-response function where 3 metabolites are important, responses are independent. This scenario is most favorable for BVCKMR:
    \begin{align}
        \boldsymbol{g}(X, t) &= \beta_1 X_1^2 - \beta_2 X_6^2 + 0.5\beta_3 X_1 X_2 + \beta_4 X_7 + \beta_5 X_8 + 
        0.3 (\beta_6 X_1^2 + \beta_7 X_7 + \beta_8 X_8) t\\
        \beta_{lj} &\sim Unif(0.25, 0.5) \bigcup Unif(-0.5, 0.25), \quad l=1,...,8 \\
        C_\xi &= I, \quad C_\epsilon = 0.5 I
    \end{align}
\end{enumerate}

We choose $K = K^\ast + 2$ for PCA and our method, and $H = 2$ for our method, to resemble analyses where $K, H$ are close to but not exactly the true latent dimension. For predictive performance evaluation, we calculate the mean predictive square error (MPSE) on a test set of $200$ subjects at 10 time points. Additionally, to evaluate how well the methods measure the relative importance of each chemical, we calculate the Spearman correlation between the true relative importance rank and the inferred rank of chemicals in $X$. The rank is based on the absolute value of the effect at each time point, summed over all time points. We calculate the rank for our model by first calculating the induced effects in the original predictors $X$ at each time point as shown in \cite{ferrari2021bayesian}. Similarly, we can calculate the effects from the PCA-LMM analysis in $X$ as $\hat{\beta}_X = V^{(K)} \hat{\beta}_{PCs}$, where $V^{(K)}$ is the first K left singular vectors, and $\hat{\beta}_{PCs}$ is a vector of regression coefficients for the PCs. Table~\ref{tab:sim-res-mpse} shows the MPSE results. Note that PCA-LMM performed worst in all scenarios, even when all its assumptions were met. Our method performs best when the data are generated according to its model. When the data are more favorable to the other models, our method still performs better than PCA-LMM. Table~\ref{tab:sim-res-rank} shows the Spearman correlation results. Here, our method performs best in the first two scenarios and is very close to the best in the third scenario.

\begin{table}[H] 
\caption{Average MPSE of 100 simulations per scenario.\label{tab:sim-res-mpse}}
\begin{tabularx}{\textwidth}{CCCC}
\toprule
\textbf{Model}	& \textbf{Scenario 1}	& \textbf{Scenario 2} & \textbf{Scenario 3}\\
\midrule
Oracle & 1.51 (0.05)          & 1.51 (0.08)          & 1.30 (0.02)\\ 
Mean predictor & 12.0 (2.65)  & 7.57 (4.6)           & 3.93 (0.25)\\ 
PCA-LMM & 7.14 (1.95)        & 5.93 (2.81)          & 3.23 (0.28)\\
BVCKMR & \textbf{4.44} (1.17) & 5.69 (2.55)          & \textbf{1.60} (0.04)\\
BMFR (our model) & 5.33 (1.46)       & \textbf{2.92} (1.31) & 2.63 (0.22)\\
\bottomrule
\end{tabularx}
\end{table}

\begin{table}[H] 
\caption{Average Spearman's correlation of rank of variable importance of 100 simulations per scenario.\label{tab:sim-res-rank}}
\begin{tabularx}{\textwidth}{CCCC}
\toprule
\textbf{Model}	& \textbf{Scenario 1}	& \textbf{Scenario 2} & \textbf{Scenario 3}\\
\midrule
     Oracle & 1              & 1 & 1\\ 
     Mean predictor & -      &  - & -\\ 
     PCA-LMM & 0.81 (0.07)  & 0.63 (0.22) & 0.29 (0.14)\\
     BVCKMR & 0.76 (0.08)    & 0.54 (0.19) & \textbf{0.59} (0.1)\\
     BMFR (our model) & \textbf{0.89} (0.06) & \textbf{0.89} (0.08) & 0.55 (0.17)\\
\bottomrule
\end{tabularx}
\end{table}

\section{Analysis of MSCEHS cohort data}
\label{sec:famr_sinai}

\subsection{Data preprocessing}

\begin{figure}[H]
    \includegraphics[width=0.5\textwidth]{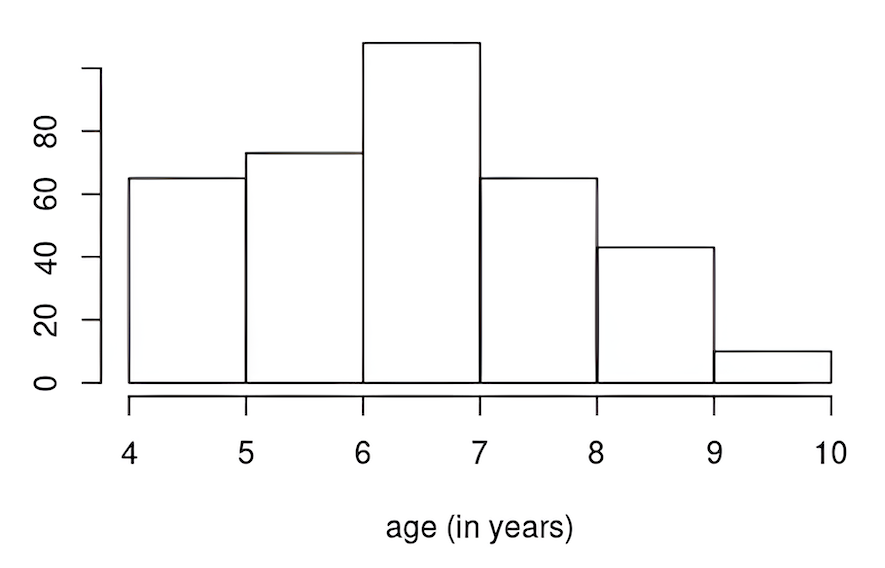}
    \caption{Histogram of ages at which the children had adiposity outcomes' measurements.}
    \label{fig:hist-age}
\end{figure}

We log-transformed WC so that its marginal is more approximately normal. We mean-centered all outcomes. We also used the logarithm of phthalate metabolites as exposures. We corrected for urinary dilution by dividing the metabolite concentrations (not on the log scale) by the Cratio as discussed in Section~\ref{sec:famr_data}. We standardize other continuous covariates and create dummy variables for categorical covariates. We used \texttt{R} package \texttt{mice} for multiple imputation of missing values in the covariates (using predictive mean matching for continuous, logistic regression for binary, and proportional odds model for ordered categorical covariates). We created a time variable that is age in years based on age in months of the children at follow-up visits. Age ranges from 4 to 10 years old (Figure~\ref{fig:hist-age}).

\unskip

\subsection{Preliminary analysis}

As a preliminary analysis, we did two-staged PCA-LMM analyses of each of the four outcomes. We applied PCA to the standardized log chemical exposures. We used the first three principal components (PCs) for the LMM stage because they explained over 90\% of the variations in the exposures. The first three PCs can be interpreted as the non-DEHP (excluding MEP) factor, DEHP factor, and MEP factor respectively. The PCA factor loading matrix is available in Appendix~\ref{chap:famr_appendix}, Section~\ref{sec:famr_xtra_plots}.

We fitted four independent LMMs for four outcomes with random intercepts and interactions between the PCs and the child's gender, controlling for all baseline covariates. We fitted a second set of LMMs with interactions between the PCs, child's gender, and child's age, and a third set with interactions between the PCs, child's gender, and polynomials of degree 2 of child's age. We used AIC, BIC, likelihood ratio tests, and 6-fold cross-validated MPSE for model comparison within each outcome. After Bonferroni correction for multiple testing, we still saw evidence from the likelihood ratio tests that models with linear interactions with age were the best fits for FMP, WHR, and WC. Models with linear interactions with age had the best cross-validated MPSE for BMIz, FMP, and WC. We saw insufficient evidence of the quadratic interactions in age being useful.
Full summary tables of AIC, BIC, p-values, and MPSEs can be found in Appendix~\ref{chap:famr_appendix}, Section~\ref{sec:famr_xtra_plots}.

\subsection{Main analysis via BMFR}

In the main analysis, we fitted our proposed Bayesian multivariate factor regression with time-varying effects to all four outcomes simultaneously. We fitted two models, one for male and one for female children, to assess any sex-specific effects. We controlled for linear main effects of covariates and included random intercepts.

We used the following prior specifications. Since we standardized the log chemical exposures to have unit variances, we set hyperparameters for inverse gamma priors on idiosyncratic noise variances $\sigma^2_{X,1},...,\sigma^2_{X,1}$ so that they are less than 1 with 99\% probability. For hyperparameters on the MGP prior for loadings $\Theta$ and $\Lambda$, we used $a_1=2.1$ and $a_2 = 3.1$ as suggested in the note by \cite{durante2017note}. For the IW prior on $\Sigma_Y$, we set $S_0$ to the sample covariance of the four outcomes, and a small $s_0 = 6$ so that prior is loosely centered around the sample covariance. For a weakly informative prior that $\nu^2$ should be below 100, we use a half-Cauchy with a scale of 25 as done in \cite{gelman2006prior}.

We selected the number of factors $K$ and the number of basis functions $H$ using grid search. We chose the combination of $K$ and $H$ from a set of options ($K\in\{2,3,4\}$, and $H\in\{1,2\}$) that had the smallest 6-fold cross-validated MPSE, with $H \leq K$ only. We considered small values for $H$ because of our prior belief that the effects of the exposures on all outcomes were similar. The final model for males had $K=3, H=1$; and the one for females had $K=3, H=2$. We fixed the length scale parameter $\kappa = 6$, since we observed from our preliminary analysis that there were likely no highly variable effects. Sensitivity analysis was also done for models that infer the length scale parameter during MCMC.

\subsection{Results of analysis via BMFR}

For out-of-sample predictive performance comparison, we calculated MPSE of a baseline mean model that returned outcome-specific means.
The 6-fold cross-validated MPSE of our proposed method is 0.93 compared to 0.98 of the baseline mean model. For the female model, our proposed method's MPSE is 0.82 compared to 1.04 of the baseline mean model. On the whole data set (combining male and female), our MPSE is 0.88, compared to 0.92 of PCA-LMMs with linear interactions with age, and 1.03 of the baseline mean model.

\begin{figure}[H]
    \includegraphics[trim={30cm 90cm 0 30cm}, clip, width=0.95\textwidth]{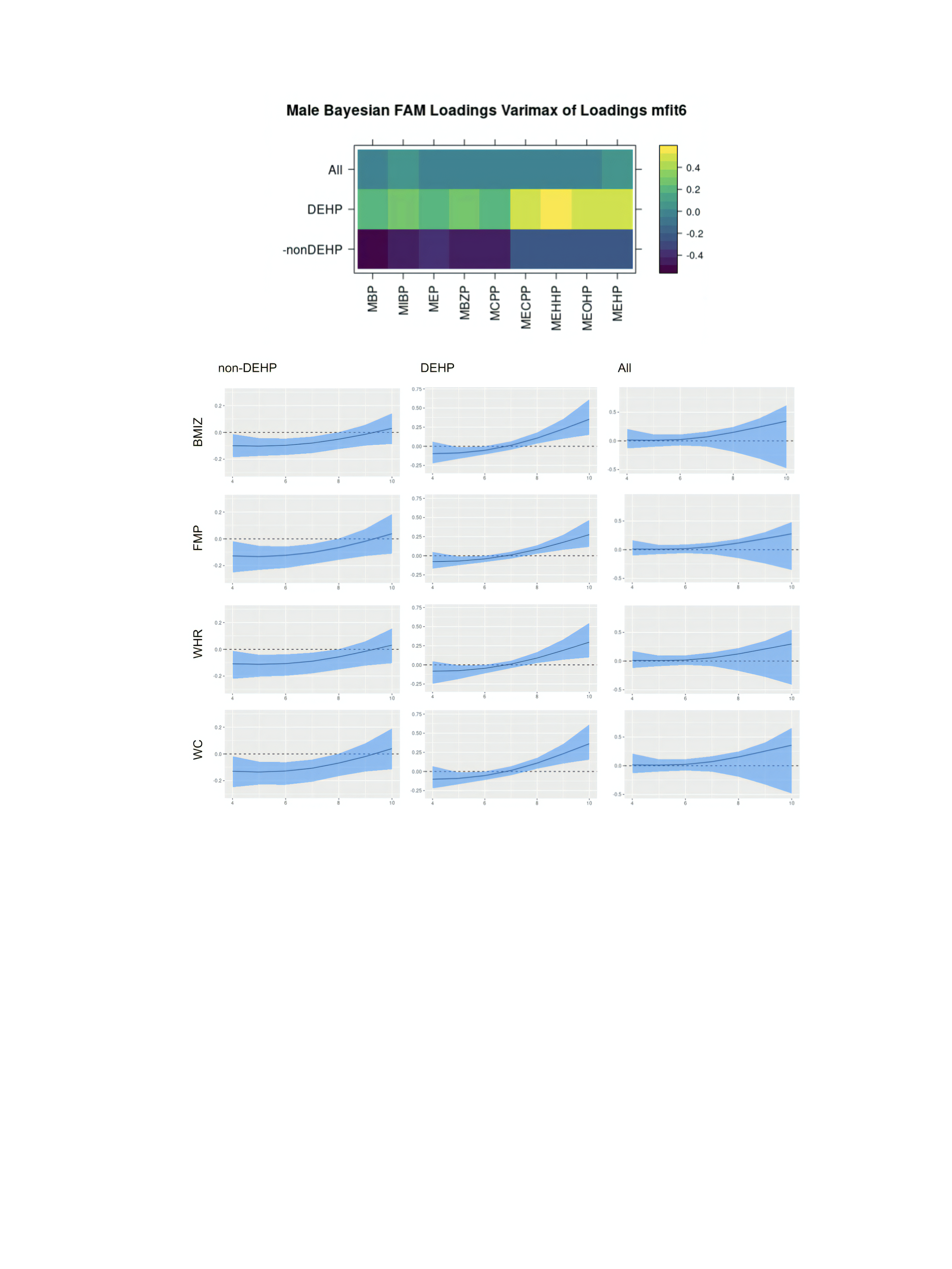}
    \caption[MatchAligned factor loading matrix and time-varying effects of each latent factor from the model fitted for males.]{MatchAligned factor loading matrix and time-varying effects of each latent factor from the model fitted for males. The blue band displays 95\% posterior credible interval, and the black solid line shows the posterior mean.}
    \label{fig:famr_male_res}
\end{figure}

\begin{figure}
    \includegraphics[trim={30cm 100cm 0 30cm}, clip, width=0.95\textwidth]{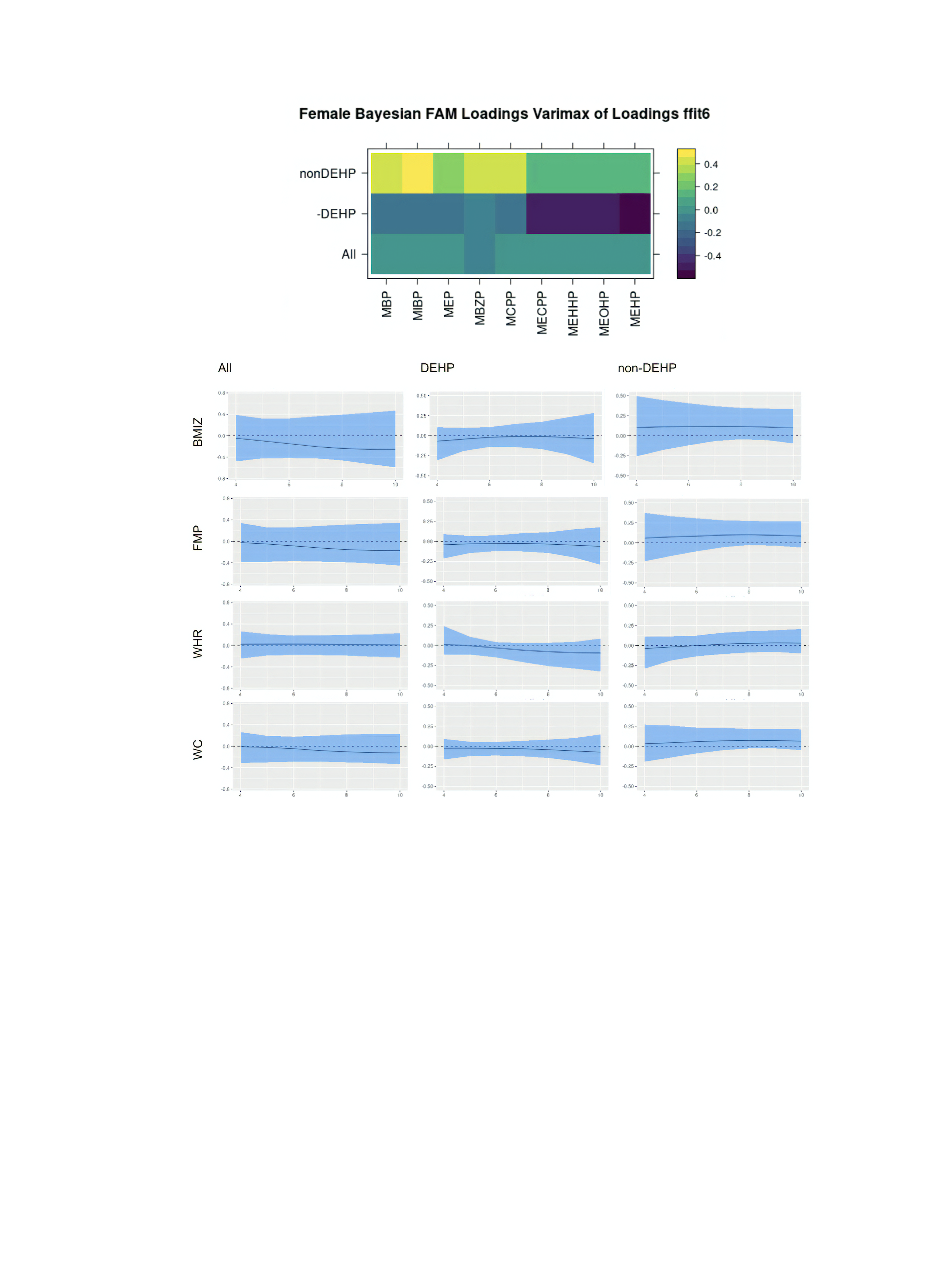}
    \caption[MatchAligned factor loading matrix and time-varying effects of each latent factor from the model fitted for females.]{MatchAligned factor loading matrix and time-varying effects of each latent factor from the model fitted for females. The blue band displays 95\% posterior credible interval, and the black solid line shows the posterior mean.}
    \label{fig:famr_female_res}
\end{figure}

For interpretation of the results, we resolved rotational and label-switching ambiguity in the factors using the MatchAlign algorithm proposed by \cite{poworoznek2021efficiently}. Figure~\ref{fig:famr_male_res} shows the post-processed factor loading matrix and time-varying effects of each latent factor from the model fitted for male children. Time-varying effects were transformed to represent the effects of one unit increase in the latent factors. We identified two main latent factors corresponding to the non-DEHP group and DEHP group of chemicals. This is consistent with previous studies on phthalate \citep{maresca2016prenatal}. Though both groups of chemicals seem to be associated with lower values in adiposity outcomes at younger ages, their promotive effects of obesity seem to increase over time. The latent factors identified in the model for female also includes non-DEHP and DEHP groups, though all effects seem to remain null over time (Figure~\ref{fig:famr_female_res}). The sex-specific effects here could be related to phthalates being anti-androgens \citep{fisher2004environmental}. Though the C.I.s from the female model are not statistically significant, the effects of latent factors on WHR seem to be different from their effects on the other three outcomes. This could be related to the fact that WHR is less correlated to the other outcomes than they are to each other.

\unskip

\unskip

We included results from sensitivity analysis of inferring the length scale parameter during MCMC in Appendix~\ref{chap:famr_appendix}, Section~\ref{sec:famr_xtra_plots}. Overall, the sensitivity analysis results were similar to results in Figure~\ref{fig:famr_male_res} and Figure~\ref{fig:famr_female_res}.

\section{Discussion}
\label{sec:famr_discussion}

This paper assesses the time-varying health effects of prenatal phthalate exposures on adiposity outcomes measured in children from ages 4 to 9 from the MSCEHS cohort study with using the BMFR approach we propose. BMFR represents phthalate mixtures as latent factors–a DEHP and a non-DEHP factor, borrows information across highly correlated adiposity outcomes to improve estimation precision, models potentially non-linear time-varying effects of the latent factors on adiposity outcomes, and fully quantifies uncertainty using state-of-the-art prior specifications. We find that in boys, at younger ages (4–6 years), all phthalate latent factors (DEHP and non-DEHP) show negative associations with adiposity outcomes. After age 7, these associations begin to become positive. In girls, there is no evidence of associations between phthalate components and outcomes. We also find these time-varying effects to be similar across all adiposity outcomes (BMIz, fat mass percentage, waist-to-hip ratio, and waist circumference). Our introduction of a new Bayesian mixture method for estimating time-varying effects with full uncertainty quantification and our finding of sex-specific time-varying associations of prenatal phthalate exposures with childhood obesity between age 4 to 9 are novel.

We were able to estimate sex-specific time-varying effects not previously identified in analyses of the MSCEHS cohort because BMFR is customized to this research question and includes several custom specifications for the data. BMFR applies prior specifications to infer modeling decisions and avoid the artificially narrow confidence intervals that are an unintended consequence of fixing choices in traditional statistical models or multistage analyses. The model includes variable selection priors for covariates, avoiding fixed subsets; Gaussian process priors for flexible modeling of time-varying effects without assuming a functional form; and half-t priors for robust between-subject variances. BMFR represents highly correlated exposures with a small number of latent factors that are independent of each other but correlated with the outcomes, while fully quantifying uncertainty in the number of components through the MGP prior. BMFR also jointly models multiple outcomes of adiposity—BMIz, waist circumference, waist-hip ratio, and fat mass percentage—available in our cohort to improve estimation precision. Finally, BMFR quantifies the uncertainty in imputing missing data and values below the limit of detection (LOD) within the MCMC sampling, propagating this uncertainty into the credible intervals for the effects of interest. An R package for BMFR is freely available at https://github.com/phuchonguyen/famr. The package implements the MCMC algorithm in Appendix \ref{sec:famr_mcmc} in C\texttt{++} for optimal computational speed.

Our finding of time-varying effects is similar to results from \citep{sol2025fetal}, which reported that exposure to phthalate in the first trimester was associated with lower BMI six months after birth but higher BMI in older ages, although they did not observe specific sex effects. Other studies have also found somewhat similar time-varying effect: higher maternal urinary phthalate concentrations associated with lower fetal growth and birth weight, followed by higher growth trajectories later on \citep{ li2021trimester, ferguson2022prenatal}, with sex-specific effects \citep{li2021trimester}. Our results also align with \citep{maresca2016prenatal}, which found non-DEHP metabolites associated with lower BMI, fat mass, and waist circumference in boys aged 5 and 7, and with \citep{montes2022modification}, which reported DEHP associated with higher outcomes among boys between ages 8 and 10, and provide a potential explanation for seemingly inconsistent results between the two studies. The time-varying effect we identify in boys may appear null when aggregating across all ages, which is consistent with a previous analysis of this cohort \citep{buckley2016nyc} that estimated average effects over time and did not observe associations with percent fat mass or sex-specific modification. However, there are other studies inconsistent with ours, including reports of DEHP associated with a lower BMI in girls \citep{buckley2016prenatal} and a higher weight gain in early childhood that stabilized during puberty in girls \citep{heggeseth2019heterogeneity}. Our findings may have important implications for pregnancy care guidelines and child health. The time-varying effects of prenatal exposure suggest that these exposures may influence childhood obesity years after the time of exposure. They further suggest that the third trimester of pregnancy may be a vulnerable window of exposure, making interventions to reduce phthalate exposure during pregnancy important. 

This study also has limitations and opportunities for future improvements. BMFR does not model nonlinear dose-response relationships. As a result, we did not investigate different effects at different exposure levels, though this could be done by stratifying the analysis by tertiles of exposure. This may be important, as mixture effects could vary in nonlinear ways across exposure levels \citep{bobb2015bayesian, liu2018bayesian}. In this cohort, exposures were measured from a single urine sample, with collection times ranging from 25 to 40 weeks of gestation. Given the short half-lives of urinary phthalate metabolites \citep{samandar2009temporal} and the likely episodic nature of exposures \citep{buckley2016nyc}, this limits the precision of exposure measurement. Future research should consider multiple exposure measurements and aim to identify the most critical window of vulnerability during the fetal period to growth trajectories. Data on adiposity outcomes before age 4 and after age 9 are not available in this cohort. Stratification by sex also reduces sample size and power to detect small effects in our analysis. Our findings would be strengthened by replication across a longer time window, from birth through puberty, and with data from a larger cohort study. Finally, residual confounding may remain, as we could not account for child calorie intake, or maternal consumer product preferences, which could influence exposure levels \citep{buckley2016nyc}.

\section{Conclusions}

This paper presents a Bayesian multivariate factor regression approach to assessing the time-varying health effects of prenatal phthalate exposures as measured in maternal urine sample during the third trimester of pregnancy on adiposity outcomes measured in young children from age 4 to 9 using data from the MSCEHS cohort study. BMFR addresses challenges in analyzing mixture exposures by representing them as latent factors that predict the outcomes. It also allows non-linear time-varying effects of exposure mixture to be estimated with full uncertainty quantification, while improving estimate precision by borrowing information across correlated adiposity outcomes. The results show in boys, at younger ages (4-6) all phthalate components show association with lower adiposity outcomes, however, after age 7, they begin to show association with higher outcomes. In girls, there is no evidence of associations phthalate components and adiposity outcomes.



\vspace{6pt} 





\authorcontributions{Conceptualization, all authors; methodology, Phuc H. Nguyen and Amy H. Herring; implementation, formal analysis, and validation, Phuc H. Nguyen; data curation, Amy H. Herring and Stephanie M. Engel; writing---original draft preparation, Phuc H. Nguyen; writing---review and editing, all authors; visualization, Phuc H. Nguyen; supervision, Amy H. Herring; funding acquisition, Amy H. Herring. All authors have read and agreed to the published version of the manuscript.}

\funding{Phuc H. Nguyen and Amy H. Herring were funded by grants R01ES027498 and R01ES028804 of the National Institute of Environmental Health Sciences of the United States National Institutes of Health. Stephanie M. Engel was partially supported by research funding from grants R01ES035625, P30ES010126, RD-84021901, R01ES033518, and R01ES027498.}

\institutionalreview{The study was secondary analysis of de-identified data from the Mount Sinai Children’s Environmental Health Study, and has received Administrative Review from the ethics committee of the the University of North Carolina at Chapel Hill Office of Human Research Ethics (study number 11-0317, Reference ID 447995, and annual administrative review acknowledgment date of 10/14/2024).}

\informedconsent{Not applicable.}

\dataavailability{The data presented in this study are available on request.}

\acknowledgments{We want to thank David Dunson, Joseph Mathews, Youngsoo Baek, and Raphaël Morsomme for the helpful discussions about this work.}

\conflictsofinterest{Author Phuc H. Nguyen was employed by the company LinkedIn Corporation. The remaining authors declare that the research was conducted in the absence of any commercial or financial relationships that could be construed as a potential conflict of interest.} 



\abbreviations{Abbreviations}{
The following abbreviations are used in this manuscript:
\\

\noindent 
\begin{tabular}{@{}ll}
NHANES & National Health and Nutrition Examination Surveys\\
MSCEHS & Mount Sinai Children’s Environmental Health Study\\
DEHP & di-2-ethylhexyl phthalate \\
LOD & limits of detection\\
BMI & body mass index\\
FMP & fat mass percentage\\
WHR & waist-to-hip ratio\\
WC & waist circumference\\
BVCKMR & Bayesian varying coefficient kernel machine regression\\
BMFR & Bayesian multivariate factor regression\\
PCA & principal component analysis\\
\end{tabular}
}

\appendixtitles{no}\label{chap:famr_appendix} 
\appendixstart
\appendix
\section[\appendixname~\thesection]{MCMC Algorithm}\label{sec:famr_mcmc}
Let $X$ be an $n \times p$ matrix of mixtures data, $Y_{it}$ be a vector of $q$ outcomes at time $t$ for subject $i$,  $K$ be a bound on the number of latent factors,

\begin{enumerate}
    \item Sample idiosyncratic noise variances for the mixtures $\Sigma_X$:
    \begin{align}
        \sigma^{-2}_{X,j}|. &\sim G(\frac{2.5 + n}{2},
        \frac{2.5(0.084) + \sum_{i=1}^n(X_{ij}-\theta_j^T\eta_i)^2}{2})
    \end{align}
    
    where $\theta_j$ is the $j^{th}$ row of loading matrix $\Theta$, for $j=1,...,p$. We use prior $\sigma^{-2}_{X,j} \sim G(\frac{2.5}{2}, \frac{2.5(0.084)}{2})$ since columns of $X$ have been standardized to have variance 1.
    
    \item Sample the mixtures loading matrix $\Theta$:
    \begin{align}
        \theta_j | . &\sim N_K((D_j^{-1} + \frac{\eta^T\eta}{\sigma^2_{X,j}})^{-1}\frac{\eta^T X_{.j}}{\sigma^2_{X,j}}, (D_j^{-1} + \frac{\eta^T\eta}{\sigma^2_{X,j}})^{-1}) \\
    \end{align}
    
    where $X_{.j}$ is the $j^{th}$ column of $X$, $\theta_j$ is the $j^{th}$ row of $\Theta$.
    
    $$\phi_{jk}|. \sim G(\frac{v+1}{2}, \frac{v +\tau_k \theta_{jk}^2}{2})$$
    
    for $k=1,...,K$ and $v=3$ as suggested in \citep{bhattacharya2011sparse}.
    
    $$\delta_1|. \sim G(a_1 + \frac{pK}{2}, 
    1 + \frac{1}{2}\sum_{l=1}^K \tau_l^{(1)} \sum_{j=1}^p \phi_{jl} \theta_{jl}^2)$$

    $$\delta_k| . \sim G(a_2 + \frac{p(K-k+1)}{2}, 1 + \frac{1}{2}\sum_{l=1}^K \tau_l^{(k)} \sum_{j=1}^p \phi_{jl} \theta_{jl}^2)$$
    
    for $k \geq 2$, where $\tau_l^{(k)} = \prod_{t=1, t\neq k}^l \delta_t$, for $k=1,...,K$. 

    Set $\tau_k = \prod_{t=1}^k \delta_t$.

    We set $a_1=2.1$ and $a_2=3.1$ following the note by \citep{durante2017note}. 
    
    \item Sample latent factors for each subject using adaptive Metropolis-within-Gibbs: 
    
    Sample a proposal $\eta_i^\ast \sim N(\eta_i^{(s)}, \Tilde{s}_i^{(s)})$, where $\eta_i^{(s)}$ is the value for subject $i$ at iteration $s$.

    The log posterior density at $\eta_i$ (after integrating out $\xi_i$) is proportional to:

    \begin{align}
    l(\eta_i) &= l_{X_i}(\eta_i) + l_\eta(\eta_i) + \sum_t^{T_i} l_{Y_{it}}(\eta_i)\\
    l_{X_i}(\eta_i) &= -\frac{1}{2} \eta_i^T \Theta^T \Sigma_X^{-1} \Theta \eta_i; \quad l_\eta(\eta_i) = -\frac{1}{2}\eta_i^T \eta_i\ \\
    l_{Y_{it}}(\eta_i) &= - \frac{1}{2} (1 + \nu^2)^{-1}m_{it}(\eta_i) ^T \Sigma_Y^{-1} m_{it}(\eta_i) \\
    m_{it}(\eta_i) &= Y_{it} - (B(t) \eta_i + \sum_{l} z_{itl} B_l^{(in)} \eta_i + B^{(c)}Z_{it}) 
    \end{align}
    
    With probability $\min(1, exp\{l(\eta_i^\ast) - l(\eta_i^{(s)})\})$, accept the proposal and set $\eta_i^{(s+1)}=\eta_i^\ast$. Otherwise, set $\eta_i^{(s+1)}=\eta_i^{(s)}$.

    Update the proposal scaling $\Tilde{s}_i^{(s)}$ every 50 iterations according to Chapter 4, section 4.3 in \citep{brooks2011handbook}.
    

    \item Sample subject-level random effects:
    \begin{align}
        \xi_i |. &\sim N(
        (\frac{1}{\nu^2} + T_i)^{-1} \sum_t^{T_i} \Tilde{Y}_{it}, 
        (\frac{1}{\nu^2} + T_i)^{-1} \Sigma_Y )
    \end{align}
    
    for $i=1,...,n$, where $T_i$ is the number of follow-up times of subject $i$, and $\Tilde{Y}_{it} = Y_{it} - B(t)\eta_i - \sum_l^L z_{it l} B_l^{(in)}\eta_i - B^{(c)}Z_{it}$.

    Sample $\nu^2|.$ using Metropolis Hasting.

    \item Sample basis function loading matrix $\Lambda$:

    \begin{align}
        \lambda_{.h}|. &\sim N_q(V_{.h} m_{.h}, V_{.h}) \\
        m_{.h} &= \sum_{i,t} u_{h.}(t)^T \eta_i Y_{it}\\
        V_{.h} &= \left[ D^{-1}_h + \Sigma^{-1} (\sum_{i,t} u_{h.}(t)^T \eta_i )^2 \right]^{-1}
    \end{align}

    \noindent where $D_h = \tau^\ast_h diag(\phi^\ast_{1h},...,\phi^\ast_{qh})$ contains MGP shrinkage parameters. Sample these the same as in Step 2.

    \item Sample basis functions $U(t)$:
    \begin{align}
        D^{(hk)}_i &= \begin{bmatrix}
                    \lambda_{.h}\eta_{ik}O_{i1} & ... & 0\\
                    0 & ... & 0 \\
                    0 & ... & \lambda_{.h}\eta_{ik}O_{iT}
                    \end{bmatrix}\\
        v^{(hk)}_{it} &= \begin{bmatrix}
                    \sum_{l\neq h} \lambda_{1l}\eta_{ik}u(t)_{lk} \\
                    ...\\
                    \sum_{l\neq h} \lambda_{ql}\eta_{ik}u(t)_{lk}
                    \end{bmatrix}\\
        v^{(hk)}_{i} &= \begin{bmatrix}
            v^{(hk)}_{i1} \\ ... \\ v^{(hk)}_{iT}
        \end{bmatrix}\\
        u_{(hk)} &\sim N(V^{(hk)}m^{(hk)}, V^{(hk)}) \\
        V^{(hk)} &= \left[ C^{-1} + \sum_i (D^{(hk)}_i)^T \Sigma^{-1} D^{(hk)}_i \right]^{-1}\\
        m^{(hk)} &= \sum_i (D^{(hk)}_i)^T \Sigma^{-1}  (y_i^{(k)} - v^{(hk)}_{i})
    \end{align}

    \item Sample Gaussian Process bandwidth $\kappa$ from a set of plausible values:

    Sample one option $\kappa^\ast$ w.p. $\frac{l(\kappa^\ast; U)}{\sum l(\Tilde{\kappa}; U)}$ where $l(\kappa^\ast; U)$ is the likelihood of $\kappa^\ast$.
     
    \item Sample matrix of main effects of covariates:
    \begin{align}
        B^{(c)}|. &\sim N_{q \times L}(\frac{\Tilde{Y}^TZ}{1 + \nu^2} ( \frac{Z^TZ}{1 + \nu^2} + \Psi^{(c) -1})^{-1}, 
        \Sigma_Y,
        (\frac{Z^TZ}{1 + \nu^2} + \Psi^{(c)-1})^{-1})
    \end{align}
    where $\Tilde{Y} = [\Tilde{Y}_{11},...,\Tilde{Y}_{1T_1},...,\Tilde{Y}_{nT_n}]^T$ is an $\sum_i T_i \times q$ matrix, and $ \Tilde{Y}_{it} = Y_{it} - B(t)\eta_i - \sum_l^L z_{itl} B_l^{(in)}\eta_i$. Sample shrinkage parameters for the linear regression coefficient matrix $B^{(c)}$:
    \begin{align}
        \psi_{l}^{(c)}|. &\sim GIG(u-\frac{q}{2},2\zeta_{l}^{(c)}, B_{l}^{(c)T}\Sigma^{-1}B_{l}^{(c)}) \\
        \zeta_{l}^{(c)}|. &\sim G(v, r+\psi_{k}^{(c)})
    \end{align}
    where $B_{l}^{(c)}$ is the $l^{th}$ column of $B^{(c)}$, for $l=1,...,L$.
    
    \item Sample the outcomes' noise variance parameters: 
    \begin{align}
        \Sigma|. &\sim IW(\sum_i^n T_i+ KT + KL + L +s_0, SS + s_1I_q) \\
        SS &= 
    Y^{\dagger}Y^{\dagger T}(1 + \nu^2)^{-1} + 
    \sum_k^K B_k (\psi_kC)^{-1} B_k^T \\
    &+
    \sum_l^L B_l^{(in)} {\Psi^{(in)}}^{-1} B_l^{(in)T} + 
    B^{(c)}{\Psi^{(c)}}^{-1}{B^{(c)}}^{T}
    \end{align}
    
    where $Y^{\dagger} = [Y^{\dagger}_{11},...,Y^{\dagger}_{1T_1},...,Y^{\dagger}_{nT_n}]$ is a $q \times \sum_i^n T_i$ matrix, and 
    \newline$Y^{\dagger}_{it} = Y_{it} - B(t)\eta_i - \sum_l^L z_{itl} B_l^{(in)}\eta_i - B^{(c)}Z_{it}$ .
\end{enumerate}

We use the following parametrization of generalized inverse Gaussian \newline $y \sim GIG(p, a, b)$ if $p(y) \propto y^{p-1}e^{-(ay + b/y)/2}$ for $y>0$.

\section[\appendixname~\thesection]{Analysis of Mount Sinai birth cohort data}
\label{sec:famr_xtra_plots}
\subsection[\appendixname~\thesubsection]{Preliminary analysis}

\begin{figure}[H]
    \includegraphics[trim={0 0 0 10cm}, clip, width=0.7\textwidth]{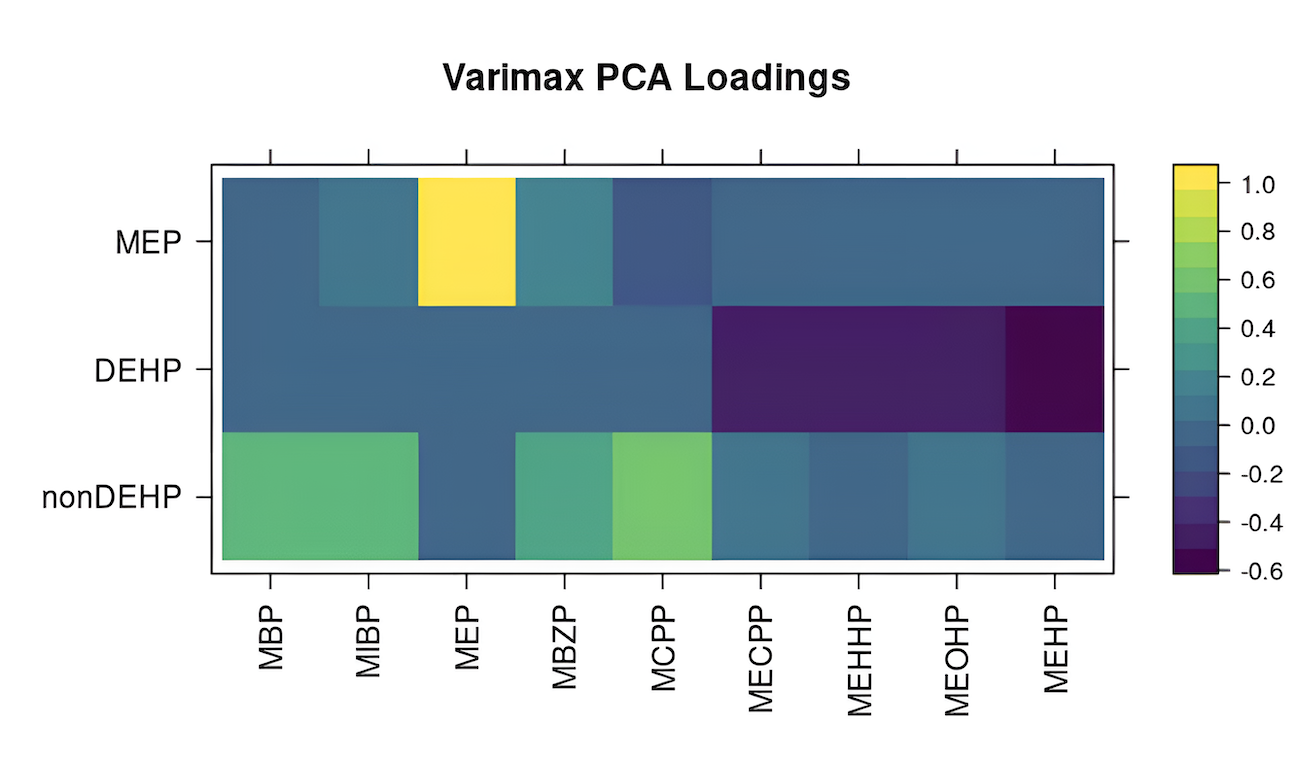}
    \caption{PCA varimax factor loading matrix of phthalate metabolites}
    \label{fig:pca-loads}
\end{figure}

\begin{table}[H]
    \caption{Model comparison for including none, linear, or polynomial degree 2 interactions with time to predict BMIz.}
    \begin{tabularx}{\textwidth}{CCCCCC}
    \toprule
    \textbf{Interaction}& \textbf{AIC}& \textbf{BIC}& \textbf{Chisq}& \textbf{p-value}& \textbf{MPSE}\\
    \midrule
     none & 727. & 809 & - & - & 1.05\\ 
     linear & 736 & 845 & 4.87 & 0.67 & 1.03\\ 
     quadratic & 747 & 883 & 3.24 & 0.86 & 1.13\\
    \bottomrule
    \end{tabularx}
    \label{tab:famr-anova-bmiz}
\end{table}

\begin{table}[H]
    \caption{Model comparison for including none, linear, or polynomial degree 2 interactions with time to predict FMP.}
    \begin{tabularx}{\textwidth}{CCCCCC}
    \toprule
    \textbf{Interaction}& \textbf{AIC}& \textbf{BIC}& \textbf{Chisq}& \textbf{p-value}& \textbf{MPSE}\\
    \midrule
     none & 859 & 941 & - & - & 1.03\\ 
     linear & 743 & 852 & 130 & $<2e-16$ & 0.90\\ 
     quadratic & 749 & 885 & 8 & 0.33 & 1.32\\
    \bottomrule
    \end{tabularx}
    \label{tab:famr-anova-fmp}
\end{table}

\begin{table}[H]
    \caption{Model comparison for including none, linear, or polynomial degree 2 interactions with time to predict WHR.}
    \begin{tabularx}{\textwidth}{CCCCCC}
    \toprule
    \textbf{Interaction}& \textbf{AIC}& \textbf{BIC}& \textbf{Chisq}& \textbf{p-value}& \textbf{MPSE}\\
    \midrule
     none & 1004 & 1086 & - & - & 0.94\\ 
     linear & 988 & 1097 & 30 & $7e-5$ & 0.98\\ 
     quadratic & 986 & 1123 & 15 & 0.03 & 2.61\\
    \bottomrule
    \end{tabularx}
    \label{tab:famr-anova-whr}
\end{table}

\begin{table}[H] 
\caption{Model comparison for including none, linear, or polynomial degree 2 interactions with time to predict WC.}
\begin{tabularx}{\textwidth}{CCCCCC}
\toprule
\textbf{Interaction}& \textbf{AIC}& \textbf{BIC}& \textbf{Chisq}& \textbf{p-value}& \textbf{MPSE}\\
\midrule
     none & 939 & 1021 & - & - & 1.1\\ 
     linear & 752 & 861 & 201 & $<2e-16$ & 0.80\\ 
     quadratic & 761 & 898 & 4.4 & 0.73 & 1.73\\
\bottomrule
\end{tabularx}
\label{tab:famr-anova-wc}
\end{table}

\subsection[\appendixname~\thesubsection]{Sensitivity analysis}
\begin{figure}[H]
    \includegraphics[trim={20cm 100cm 5cm 12cm}, clip, width=\textwidth]{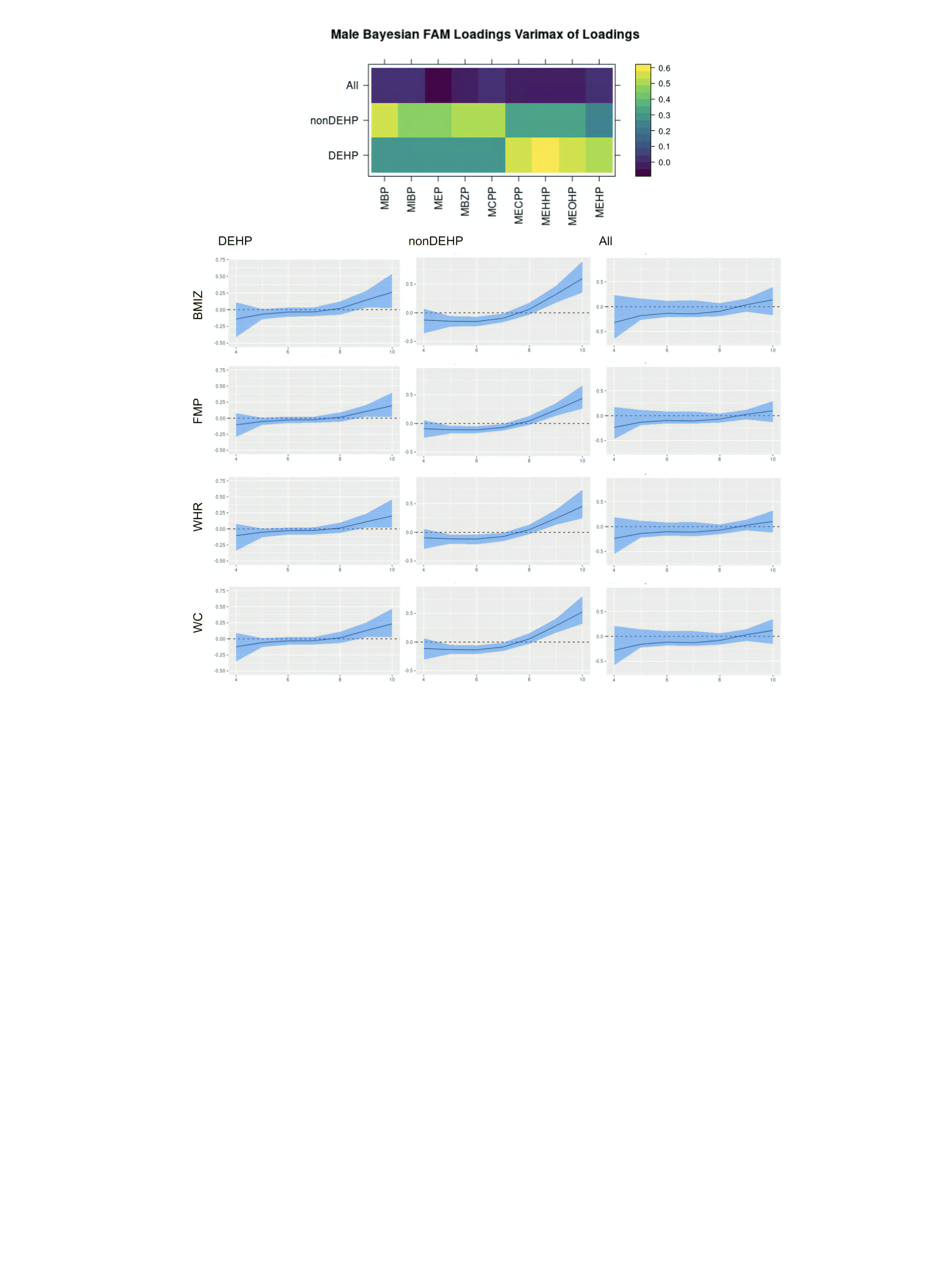}
    \caption{MatchAligned factor loading matrix and time-varying effects of each latent factor from the model fitted for male with an inferred $\kappa$.}
    \label{fig:famr_male_res_app}
\end{figure}

\begin{figure}[H]
    \includegraphics[trim={30cm 100cm 2cm 10cm}, clip, width=0.9\textwidth]{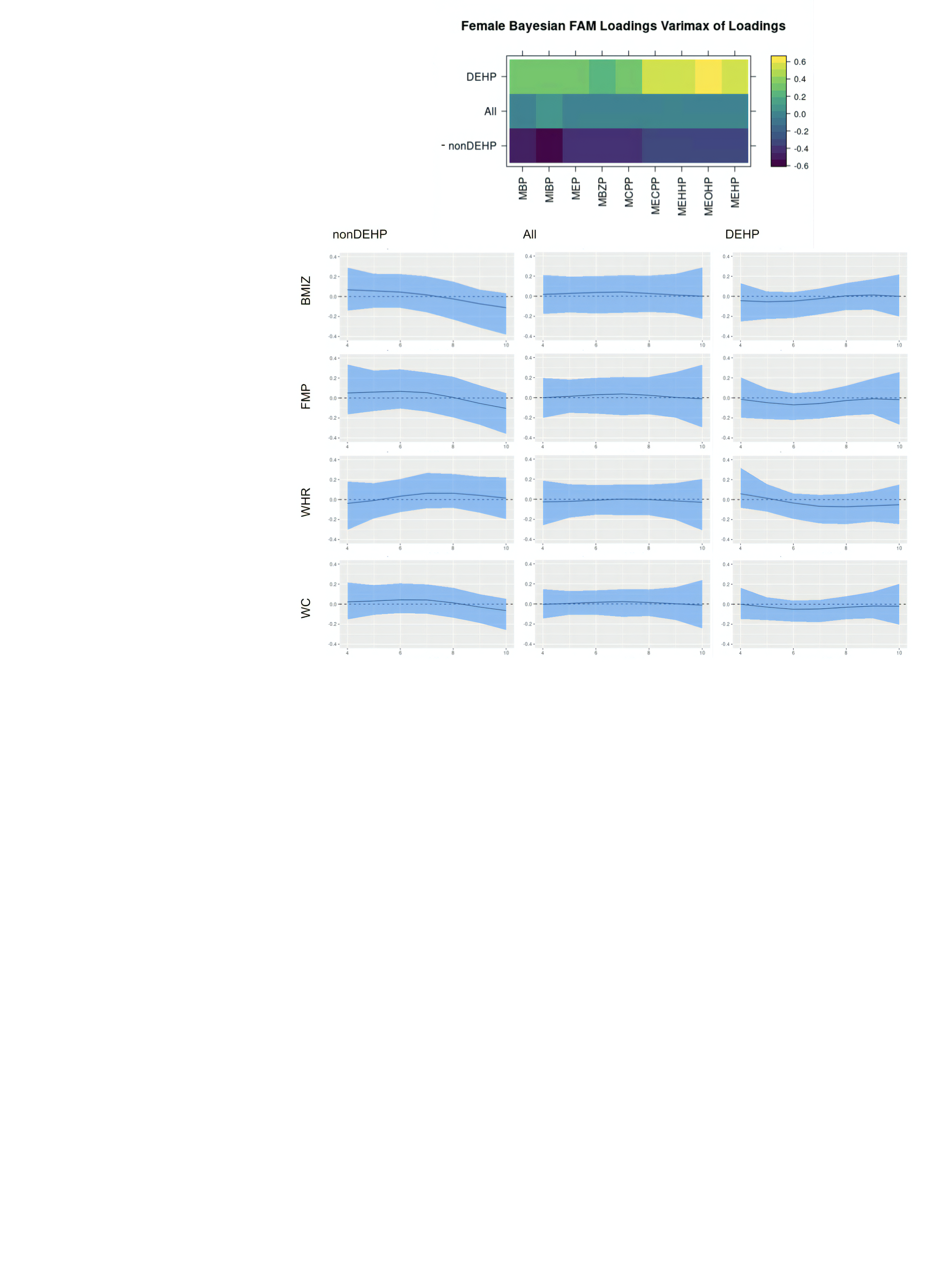}
    \caption{MatchAligned factor loading matrix and time-varying effects of each latent factor from the model fitted for females with an inferred $\kappa$.}
    \label{fig:famr_female_res_app}
\end{figure}

\isPreprints{}{
\begin{adjustwidth}{-\extralength}{0cm}
} 

\reftitle{References}
\bibliography{refs}

\PublishersNote{}
\isPreprints{}{
\end{adjustwidth}
} 
\end{document}